\documentclass[sigconf]{acmart}

\acmSubmissionID{102}
\usepackage{color}
\usepackage{bbm}
\usepackage{multirow}
\usepackage[inline]{enumitem}
\usepackage{sistyle}
\usepackage[ruled]{algorithm2e}
\usepackage{booktabs}
\usepackage{caption}
\SIthousandsep{,}
\usepackage{makecell}
\usepackage[skip=0pt]{caption}
\usepackage{amsmath}
\usepackage{hyperref}
\usepackage{array} 
\usepackage{graphicx}
\usepackage{xcolor}
\usepackage{acronym}
\usepackage[export]{adjustbox}
\usepackage{subcaption}

\newcommand{\heading}[1]{\vspace*{0.5mm}\noindent\textbf{#1.}}

\AtBeginDocument{%
  \providecommand\BibTeX{{%
    \normalfont B\kern-0.5em{\scshape i\kern-0.25em b}\kern-0.8em\TeX}}}

\makeatletter
\g@addto@macro\normalsize{%
  \abovedisplayskip 2pt plus1pt 
  \belowdisplayskip 2pt plus1pt
  \abovedisplayshortskip  2pt plus1pt%
  \belowdisplayshortskip  1pt plus1pt
}
\makeatother

\newcommand{\Direct}{Direct QA\xspace}
\newcommand{\RAG}{Vanilla RAG\xspace}
\newcommand{\Ins}{Instruction Injection\xspace}
\newcommand{\Both}{Passage Injection\xspace}

\acrodef{IR}{information retrieval}
\acrodef{LLM}{large language model}

\AtBeginDocument{%
  \providecommand\BibTeX{{%
    Bib\TeX}}}

\copyrightyear{2025}
\acmYear{2025}
\setcopyright{cc}
\setcctype{by}
\acmConference[SIGIR-AP 2025]{Proceedings of the 2025 Annual International ACM SIGIR Conference on Research and Development in Information Retrieval in the Asia Pacific Region}{December 7--10, 2025}{Xi'an, China}
\acmBooktitle{Proceedings of the 2025 Annual International ACM SIGIR Conference on Research and Development in Information Retrieval in the Asia Pacific Region (SIGIR-AP 2025), December 7--10, 2025, Xi'an, China}\acmDOI{10.1145/3767695.3769515}
\acmISBN{979-8-4007-2218-9/2025/12}



\author{Minghao Tang}
\orcid{0009-0002-1911-5142}
\affiliation{
    \institution{State Key Laboratory of AI Safety,}
    \institution{Institute of Computing Technology,}
    \institution{Chinese Academy of Sciences}
	\institution{University of Chinese Academy of Sciences}
	\city{Beijing}
	\country{China}
}
\email{tangminghao25@mails.ucas.ac.cn}

\author{Shiyu Ni}
\orcid{0009-0001-7965-7771}
\affiliation{
    \institution{State Key Laboratory of AI Safety,}
    \institution{Institute of Computing Technology,}
    \institution{Chinese Academy of Sciences}
	\institution{University of Chinese Academy of Sciences}
	\city{Beijing}
	\country{China}
}
\email{nishiyu23z@ict.ac.cn}

\author{Jiafeng Guo}
\orcid{0000-0002-9509-8674}
\affiliation{
    \institution{State Key Laboratory of AI Safety,}
    \institution{Institute of Computing Technology,}
    \institution{Chinese Academy of Sciences}
	\institution{University of Chinese Academy of Sciences}
	\city{Beijing}
	\country{China}
}
\email{guojiafeng@ict.ac.cn}

\author{Keping Bi}
\orcid{0000-0001-5123-4999}
\authornote{Keping Bi is the corresponding author.}
\affiliation{
    \institution{State Key Laboratory of AI Safety,}
    \institution{Institute of Computing Technology,}
    \institution{Chinese Academy of Sciences}
	\institution{University of Chinese Academy of Sciences}
	\city{Beijing}
	\country{China}
}
\email{bikeping@ict.ac.cn}

\renewcommand{\shortauthors}{Minghao Tang et al.}

\begin{document}
\title{Injecting External Knowledge into the Reasoning Process Enhances Retrieval-Augmented Generation}



\begin{abstract}
Retrieval-augmented generation (RAG) has been widely adopted to augment large language models (LLMs) with external knowledge for knowledge-intensive tasks. However, its effectiveness is often undermined by the presence of noisy (i.e., low-quality) retrieved passages. Enhancing LLMs' robustness to such noise is critical for improving the reliability of RAG systems.
Recent advances have equipped LLMs with strong reasoning and self-reflection capabilities, allowing them to identify and correct errors in their reasoning process.
Inspired by this ability, we propose Passage Injection—a simple yet effective method that explicitly incorporates retrieved passages into LLMs' reasoning process, aiming to enhance the model's ability to recognize and resist noisy passages. 
We validate Passage Injection under general RAG settings using BM25 as the retriever. Experiments on four reasoning-enhanced LLMs across four factual QA datasets demonstrate that Passage Injection significantly improves overall RAG performance.
Further analysis on two noisy retrieval settings-random noise, where the model is provided irrelevant passages, and counterfactual noise, where it is given misleading passages-shows that Passage Injection consistently improves robustness.
Controlled experiments confirm that Passage Injection can also effectively leverage helpful passages. These findings suggest that incorporating passages in LLMs' reasoning process is a promising direction for building more robust RAG systems. \looseness=-1
\footnote{The code and data can be found at: \url{https://github.com/Trustworthy-Information-Access/Passage-Injection}}
\end{abstract}

\maketitle

\begin{figure*}[ht]
    \centering
    \includegraphics[width=\linewidth]{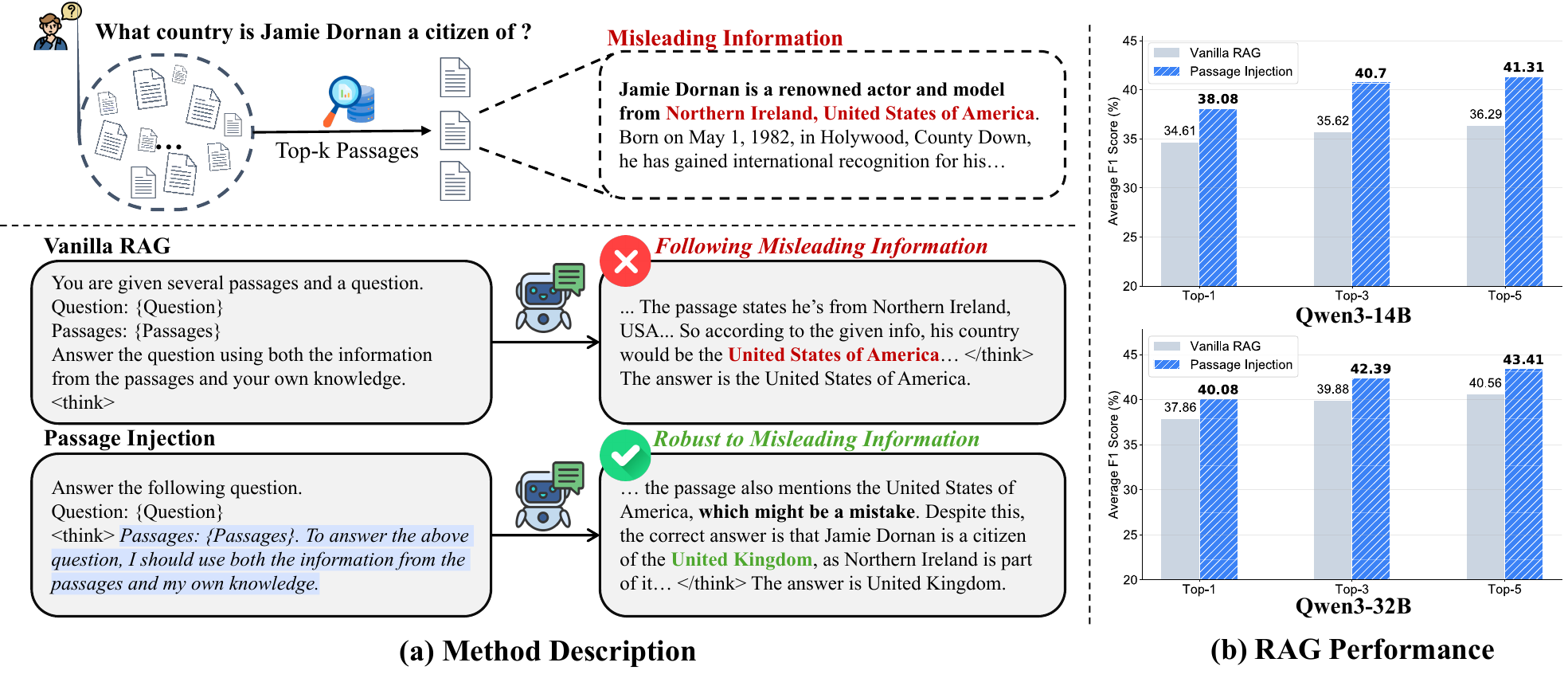}
    \caption{(a) An example where the retrieved passages contain misleading information. The passage incorrectly states that Northern Ireland is part of the United States, while the correct answer is the United Kingdom. In this case, Vanilla RAG mistakenly follows the external misleading information and produces an incorrect answer. In contrast, Passage Injection identifies the external misinformation and generates the correct answer, demonstrating its enhanced robustness to noisy passages. (b) The performance of Qwen3-14B and Qwen3-32B under general RAG settings with different methods. We use BM25 to retrieve documents and provide the top 1, 3, and 5 most relevant documents to the model. The results show that Passage Injection significantly improves RAG performance across different numbers of passages.}
    \label{fig:Method}
\end{figure*}

\section{Introduction}

Retrieval-augmented generation (RAG) has emerged as an effective approach to improving the performance of large language models (LLMs) on knowledge-intensive tasks \cite{lewis2020retrieval, zamani2022retrieval, ram2023context}. However, the quality of retrieved documents cannot be guaranteed where low-quality documents may mislead the model, resulting in incorrect answers \cite{shi2023large, ni2024llms}. Enhancing LLMs’ robustness to such noisy passages is crucial for improving the reliability of RAG systems.

Recent advances have endowed LLMs with strong reasoning and self-reflection capabilities (i.e., reasoning-enhanced LLMs) \cite{jaech2024openai,guo2025deepseek,yang2025qwen3}. These models generate intermediate reasoning steps and leverage self-reflection to detect and correct errors in their reasoning before producing final answers.
Inspired by this ability to detect and revise reasoning errors, we speculate that if noisy passages are integrated into the model’s reasoning process, the model may better recognize the noise and remain robust.
Based on this, we propose \textbf{Passage Injection}—a method that explicitly integrates retrieved passages into the model’s reasoning process to enhance robustness against noisy information and improve overall RAG performance.

We first validate the effectiveness of Passage Injection in improving the performance under general RAG scenarios, comparing it to vanilla RAG, which places retrieved passages directly in the input prompt. We use BM25~\cite{robertson2009probabilistic}, a widely used and powerful retriever for passage retrieval.
To evaluate the impact of Passage Injection across varying question difficulties, we conduct experiments on multi-hop factual question answering (QA) datasets—2WikiMultiHopQA \cite{ho2020constructing}, HotpotQA \cite{yang2018hotpotqa}, ComplexWebQuestions \cite{talmor2018web}—as well as on the single-hop dataset PopQA \cite{mallen2023not}. 
We adopt the Qwen3 series \cite{yang2025qwen3}, a family of representative reasoning-enhanced LLMs.
Additionally, to examine the impact on models that acquire reasoning abilities through distillation, we also choose DeepSeek-R1-Distill-Qwen-32B \cite{guo2025deepseek}. 
Experimental results show that, in almost all cases, \Both significantly improves RAG performance, demonstrating its effectiveness in enhancing general RAG capabilities. We also observe that Passage Injection yields greater improvements on multi-hop questions than on single-hop questions.
Compared to Qwen3-series models, the improvement is smaller on the distillation-based model. We speculate that this is because the model's reasoning ability is learned through imitation rather than self-exploration, limiting its capacity to correct errors in the reasoning process.

To further investigate whether the benefits arise from improved robustness to noisy passages, we conduct experiments under two types of noisy retrieval scenarios:
1) \textit{Random Noise}: We use the same four datasets as those of the general RAG scenarios. For each question, the model is provided with several randomly selected passages unrelated to the question.
2) \textit{Counterfactual Noise}: A more challenging form of noise that is more likely to mislead the model. We use ConFiQA \cite{bi2024context}, where each question is paired with a counterfactual context created by replacing the correct answer in a supporting passage with a random one.
Results show that \Both consistently achieves better QA performance across almost all cases compared to vanilla RAG, demonstrating its ability in enhancing the robustness of reasoning-enhanced LLMs against noisy passages. \looseness=-1

In addition to maintaining robustness to noisy passages, effectively leveraging helpful ones is also important. To evaluate whether \Both enhances the use of helpful passages, we conduct controlled experiments, providing only gold passages containing the correct answer.
In this setting, \Both improves the utilization of helpful passages on the smaller model. However, on larger models, \Both performs similarly to vanilla RAG, suggesting that larger models already possess a strong ability to leverage helpful information provided in the prompt.
Overall, Passage Injection enhances robustness against noise while maintaining effective use of helpful passages, suggesting that incorporating passages in LLMs’ reasoning process is a promising direction for building more robust RAG systems.


\section{Related Work} 

\subsection{Retrieval-Augmented Generation} 
Retrieval-augmented generation (RAG) enhances large language models (LLMs) with a retrieval module that retrieves relevant passages from an external corpus given a query~\cite{lewis2020retrieval, zamani2022retrieval, glass2022re2g, shi2024replug, zhang2025leveraging}. 
These retrieved documents are appended to the model’s input context, enabling the LLM to access information beyond its parametric knowledge and improving performance across a wide range of tasks \cite{lewis2020retrieval, borgeaud2022improving, izacard2023atlas}. 
A range of studies has explored how to better integrate these retrieved passages into the generation process, including prompt-level augmentation~\cite{trivedi2022interleaving, wang2024rat}, embedding-level fusion~\cite{izacard2020leveraging, dong2025decoupling}, and parametric-level adaptation~\cite{su2025parametric, tan2025dynamicparametricretrievalaugmented}. 
Our work proposes a new direction by injecting retrieved passages directly into the model’s reasoning process, enabling a tighter coupling between external knowledge and step-by-step reasoning.

\subsection{Reasoning-Enhanced LLMs} 
Recent advances in reasoning-augmented models, such as OpenAI's o1~\cite{jaech2024openai}, DeepSeek-R1~\cite{guo2025deepseek}, and Qwen3~\cite{yang2025qwen3}, have significantly expanded the capabilities of LLMs.
These models explicitly generate intermediate reasoning steps before producing final answers, which enhances their ability to handle tasks requiring multi-step inference~\cite{zhou2023solving, saparov2023language, liu2025logical}.
Such step-by-step reasoning not only enhances performance but also contributes to better interpretability and transparency in model generation (see Section \ref{sec:method} for detailed generation process).
Recent studies have also begun to examine the reasoning process itself, investigating how to supervise or intervene during generation~\cite{baker2025monitoring,zhang2025reasoning,wu2025effectively}. 
\citet{wu2025effectively} observe that the attention during reasoning primarily focuses on internally generated tokens rather than the input context, and propose inserting instructions into the reasoning trajectory to better guide the model’s thinking and enhance instruction-following capabilities. 
Our work introduces a new intervention method that incorporates retrieved passages into the reasoning process, enhancing the model's ability to recognize and resist noisy information.

\section{Methodology}  
\label{sec:method}
In this section, we introduce how reasoning-enhanced LLMs generate answers with internal knowledge, how they interact with external passages under RAG, and our proposed Passage Injection.

\subsection{Reasoning-Enhanced QA with Internal Knowledge}

\heading{Direct QA} For a given question $q$, the workflow of reasoning-enhanced LLMs can be divided into three phases: \textit{Input Phase}, \textit{Reasoning Phase}, and \textit{Response Phase}. In the Input Phase, the model receives the question and encodes it. Next, in the Reasoning Phase, the model generates its reasoning and self-reflection process within the \texttt{<think></think>} tags. Finally, after the \texttt{</think>} tag, it produces the final answer $a$ as the response. This allows the model to fully leverage its internal knowledge but may also lead to overthinking—resulting in longer reasoning paths, increased computational overhead, and a higher risk of hallucination.

Currently, there are two mainstream approaches to equip LLMs with reasoning abilities. One is through reinforcement learning, allowing the model to explore reasoning paths on its own, as seen in models like Qwen3~\cite{yang2025qwen3}. The other is via distillation, such as in DeepSeek-R1-Distill-Qwen~\cite{guo2025deepseek}.

\subsection{Augmenting Reasoning-Enhanced QA with External Knowledge}
Although current LLMs possess strong reasoning abilities, there is still knowledge they do not know and thus require supplementation from external passages. In general, for a given question $q$, we typically follow a retrieve-then-read pipeline, where we first use a retriever to retrieve a set of relevant passages $D$ from an external corpus, and then provide $D$ to the model. $D$ can contain one or more passages.

\heading{Vanilla RAG} This is the standard approach which directly concatenates $D$ with $q$ and \textit{places them in the Input Phase}. This approach may lead to insufficient attention to the documents, making it difficult for the model to identify potential errors. As shown in Figure~\ref{fig:Method}(a), we present an example where the retrieved passage contains misleading content—incorrectly stating that Northern Ireland is part of the United States, whereas it is actually part of the United Kingdom. In this case, Vanilla RAG fails to recognize the misinformation and follows it, resulting in an incorrect answer.

\heading{Passage Injection} Since reasoning-enhanced LLMs can identify errors in their own reasoning process, we think that explicitly integrating passages into this process may lead the model to treat them as part of its own reasoning, thereby improving its ability to detect and correct errors and enhancing its robustness. 
Based on this, we propose Passage Injection, which \textit{injects $D$ into the Reasoning Phase} to strengthen the model's robustness against noisy passages. The example can be found in Figure~\ref{fig:Method}(a). Passage Injection correctly identifies the misinformation in the external passage, rectifies it, and ultimately produces the correct answer.

\heading{Instruction Injection} Since injecting passages into the Reasoning Phase inevitably introduces instructions on how to use the passages, we design an ablation setting where only the instruction is injected into the Reasoning Phase, while the passages themselves are placed in the Input Phase. We refer to this as Instruction Injection, which is similar to \citet{wu2025effectively}.

For models that acquire reasoning abilities through distillation, we speculate that they are better suited for RAG scenarios, as distillation itself is a form of supervised learning with long contexts—closely resembling the task format of RAG. However, since their reasoning abilities are not self-discovered but rather learned from supervision, we argue that intervening in their reasoning paths may yield less benefit.
Moreover, we think that encouraging the model to pay more attention to external passages may reduce its overthinking, thereby lowering reasoning overhead.

\section{Experimental Setup}
In this section, we introduce our experimental settings, including the datasets, models used, retriever, and evaluation metrics.
\label{section:Setup}

\begin{table*}[ht]
\caption{The overall experimental results of \Both and other RAG methods across four QA datasets, using top-5 BM25-retrieved passages. All metrics reported are F1 scores (\%). Bold numbers indicate the best performance under each model.}
\centering
\resizebox{\textwidth}{!}{
\begin{tabular}{ccccccccccc}
    \toprule
    \multicolumn{1}{c}{\multirow{2}{*}{LLM}} & \multicolumn{1}{c}{\multirow{2}{*}{RAG Method}} & \multicolumn{4}{c}{2WikiMultihopQA} & \multicolumn{2}{c}{HotpotQA} & \multicolumn{1}{c}{\multirow{2}{*}{CWQ}} & \multicolumn{1}{c}{\multirow{2}{*}{PopQA}} & \multicolumn{1}{c}{\multirow{2}{*}{\begin{tabular}[c]{@{}c@{}}Micro-\\ Average\end{tabular}}} \\ 
    \cmidrule(lr){3-6} \cmidrule(lr){7-8}
    
    \multicolumn{1}{c}{} & \multicolumn{1}{c}{} & Bridge & Comparison & Compose & Inference & Bridge & Comparison & \multicolumn{1}{c}{} & \multicolumn{1}{c}{} & \multicolumn{1}{c}{} \\ 
    \midrule
    
    \multirow{4}{*}{Qwen3-8B} & \Direct & 39.99 & 55.99 & 10.62 & 21.17 & 22.23 & 64.63 & 38.23 & 21.28 & 27.39 \\
     & \RAG & 30.71 & 47.19 & 15.57 & 21.15 & 35.42 & 64.05 & 31.66 & 32.74 & 32.45 \\
     & \Ins & 47.47 & 62.84 & 19.03 & 35.13 & 40.52 & 69.94 & 38.41 & 35.28 & 38.60 \\
     & \Both & \textbf{53.32} & \textbf{65.39} & \textbf{21.14} & \textbf{35.56} & \textbf{40.37} & \textbf{72.79} & \textbf{40.00} & \textbf{36.64} & \textbf{40.30} \\ \midrule
     
    \multirow{4}{*}{Qwen3-14B} & \Direct & 48.95 & 59.81 & 12.13 & 24.06 & 25.57 & 66.39 & \textbf{41.80} & 24.58 & 30.85 \\
     & \RAG & 39.90 & 56.29 & 19.17 & 29.12 & 39.70 & 69.55 & 34.70 & 33.91 & 36.29 \\
     & \Ins & 52.44 & 64.78 & \textbf{21.46} & \textbf{39.55} & 42.43 & 71.37 & 38.53 & 35.77 & 40.19 \\
     & \Both & \textbf{56.62} & \textbf{65.56} & 21.37 & 37.32 & \textbf{43.01} & \textbf{72.74} & 41.46 & \textbf{36.92} & \textbf{41.31} \\ 
     \midrule
     
    \multirow{4}{*}{Qwen3-32B} & \Direct & 53.24 & 56.59 & 12.74 & 25.10 & 26.93 & 69.02 & 42.02 & 24.81 & 31.45 \\
     & \RAG & 51.00 & 66.68 & 22.83 & 37.38 & 43.29 & 72.24 & 39.92 & 35.55 & 40.56 \\
     & \Ins & 49.03 & 64.05 & 22.18 & 41.01 & 42.35 & 69.09 & 39.19 & 35.14 & 39.77 \\
     & \Both & \textbf{58.63} & \textbf{70.24} & \textbf{25.18} & \textbf{45.51} & \textbf{44.96} & \textbf{75.30} & \textbf{42.91} & \textbf{37.38} & \textbf{43.41} \\ 
     \midrule
     
    \multirow{4}{*}{\begin{tabular}[c]{@{}c@{}}DeepSeek-R1- \\ Distill-Qwen-32B\end{tabular}} & \Direct & 49.36 & 58.14 & 12.39 & 25.19 & 24.91 & 66.47 & 41.73 & 23.11 & 30.17 \\
     & \RAG & 53.39 & 69.62 & 24.96 & 42.00 & 44.34 & 73.63 & 41.88 & 37.61 & 42.63 \\
     & \Ins & \textbf{55.91} & \textbf{69.67} & 26.12 & \textbf{47.40} & \textbf{45.13} & \textbf{74.53} & 44.06 & 37.57 & 43.55 \\
     & \Both & 55.75 & 69.46 & \textbf{26.48} & 46.67 & 44.43 & 74.09 & \textbf{45.02} & \textbf{38.45} & \textbf{43.84} \\ 
     \bottomrule
\end{tabular}
}
\label{tab:RAG_Performance}
\end{table*}

\subsection{General RAG Settings}
\label{sec:general rag}
To verify whether Passage Injection can improve the performance of RAG systems, we conduct experiments under general RAG settings. For each question, we retrieve the top-$k$ passages from a Wikipedia dump~\cite{ni2025knowledge} using BM25~\cite{robertson2009probabilistic}, setting $k$ to 1, 3, and 5 to assess the effect of \Both across different numbers of retrieved passages. \looseness=-1

\heading{Datasets} For more complex questions, the passages usually contain richer knowledge, making it harder to detect incorrect information. 
To evaluate the impact of Passage Injection across varying question difficulties, we conduct experiments on multi-hop factual question answering (QA) datasets—2WikiMultiHopQA \cite{ho2020constructing}, HotpotQA \cite{yang2018hotpotqa}, ComplexWebQuestions \cite{talmor2018web}—as well as on the single-hop factual QA dataset PopQA \cite{mallen2023not}.

\heading{Metrics} We evaluate answer quality using the F1 score, which is the harmonic mean of precision and recall, capturing both the correctness and completeness of the predicted answer.

\heading{LLMs} We conduct experiments using the Qwen3 \cite{yang2025qwen3} series, a state-of-the-art open-source language model family known for its strong reasoning capabilities. 
To evaluate the effectiveness of the proposed \Both method across different model scales, we experiment with Qwen3 models of 8B, 14B, and 32B parameters.  
In addition, to assess how \Both performs on models that acquire reasoning abilities through distillation, we include DeepSeek-R1-Distill-Qwen-32B \cite{guo2025deepseek}.
All models use the recommended generation settings: temperature = 0.6 and top-p = 0.95.

\heading{Baselines} We take Direct QA, Vanilla RAG, and Instruction Injection which are mentioned in Section~\ref{sec:method} as baselines.

\subsection{Controlled Settings}
In general RAG scenarios, retrieved passages typically fall into two categories: those that are helpful for answering the question, and those that are irrelevant or even misleading. To better assess the impact of \Both on both types, we design two controlled experiments detailed below.

\heading{Noise Settings}
We design two distinct noise scenarios that simulate irrelevant and misleading content in the retrieved passages:

\begin{itemize}[leftmargin=*,itemsep=0pt,topsep=0pt,parsep=0pt]
    \item \textit{Random Noise}: This setting simulates noisy passages that introduce irrelevant content. Each question is paired with three passages randomly sampled from the corpus, which are highly likely to be irrelevant to the query. We evaluate on the same four factual QA datasets as those in Section~\ref{sec:general rag}.
    
    \item \textit{Counterfactual Noise}: A more challenging form of noise that is more likely to mislead the model. We adopt the ConFiQA \cite{bi2024context} dataset, in which each question is accompanied by a misleading context that is fluent and topically relevant but factually incorrect. These contexts are constructed by replacing key entities in gold passages with randomly selected same-type distractors.
\end{itemize}

\heading{Gold Settings}
To examine whether \Both can also facilitate the use of helpful passages, we conduct experiments on 2WikiMultihopQA~\cite{ho2020constructing} and HotpotQA~\cite{yang2018hotpotqa}, where each question is paired only with its gold passages that contain the correct answer.

\section{Results and Analysis}
In this section, we first evaluate the performance under general RAG settings. We then conduct controlled experiments with noisy and gold passages to investigate the sources of performance gains.

\subsection{Performance under General RAG}
\label{subsection:RAG_Results}

Table~\ref{tab:RAG_Performance} presents the performance of different RAG methods across four QA benchmarks using top-5 retrieved passages. The main findings are as follows:
\textbf{1) \Both consistently yields the best performance.} 
Across all models, \Both achieves the highest average F1 scores, demonstrating its effectiveness in general RAG scenarios. 
Moreover, as shown in Figure~\ref{fig:Method}(b), its performance gains remain stable across different top-$k$ values.
\textbf{2) \Both brings more gains on multi-hop QA.}
Compared to its improvements on single-hop PopQA, \Both yields more significant gains on multi-hop datasets. 
This suggests that \Both is particularly effective for questions requiring complex reasoning.
\textbf{3) Incorporating passages is crucial.}
While \Ins yields moderate improvements over \RAG on most datasets, it remains less effective than \Both.
This suggests that explicitly incorporating retrieved passages into the reasoning process is more beneficial than instructions alone.
\textbf{4) Distilled models benefit more from RAG.}
Interestingly, DeepSeek-R1-Distill-Qwen-32B underperforms Qwen3-32B in the \Direct setting but surpasses it under \RAG. 
This may be due to the nature of distillation: the model is trained to follow teacher demonstrations that involve long-context reasoning, making it more aligns with the RAG format and better at utilizing retrieved content.
\textbf{5) Diminished gains on distilled models.}
While \Both improves all models, the gains are notably smaller for DeepSeek-R1-Distill-Qwen-32B. 
This may be because its reasoning ability is acquired through supervised fine-tuning from teacher models, rather than developed through its own emergent capabilities.
Consequently, the model may be less sensitive to modifications within the \texttt{<think></think>} segment.

\subsection{Performance under Noisy Passages}

\begin{figure}[ht]
    \centering
    \includegraphics[width=\linewidth]{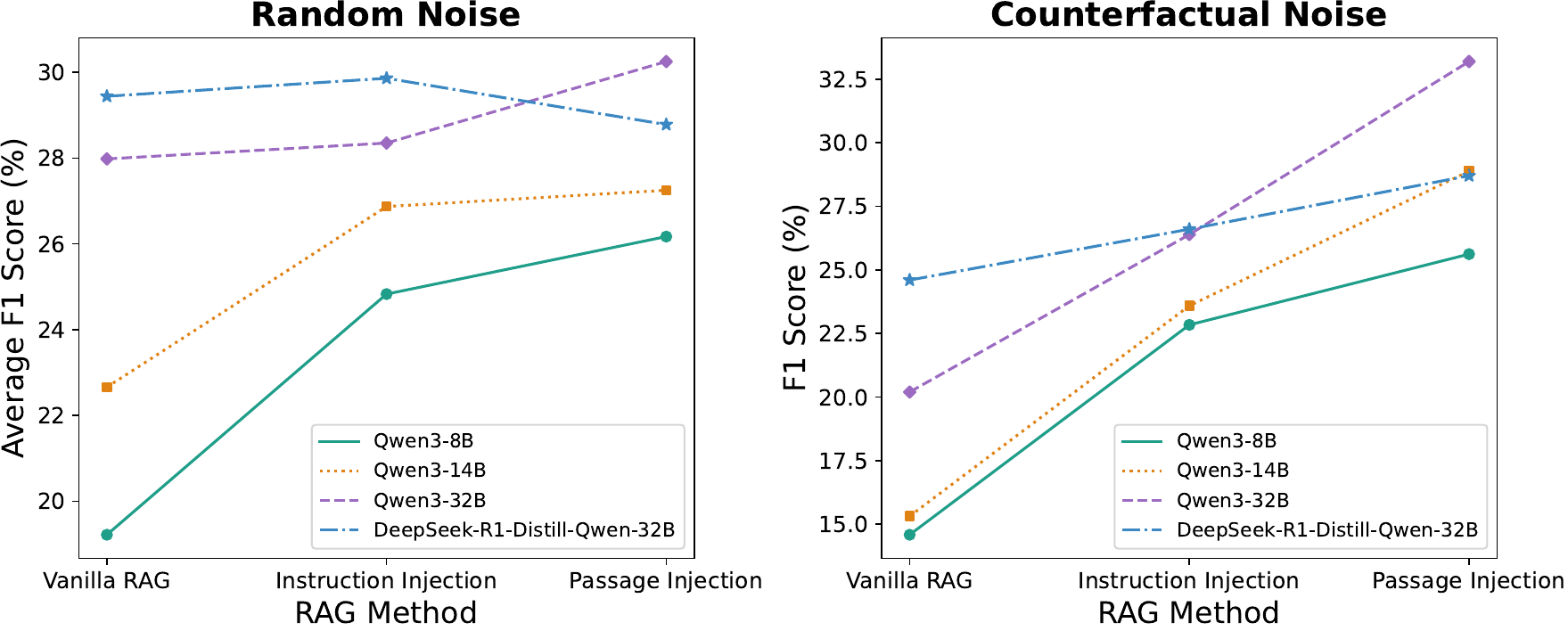}
    \caption{F1 scores under different noise settings. 
    Left: Average performance on four factual QA datasets with random unrelated passages. 
    Right: Performance on ConFiQA with counterfactual, misleading contexts.}
    \label{fig:Noise-Results}
\end{figure}

Figure~\ref{fig:Noise-Results} presents the performance of different RAG methods under both \textit{Random Noise} and \textit{Counterfactual Noise} settings. We summarize the key findings as follows:
\textbf{1) \Both consistently improves robustness.} 
Under both noise settings, \Both significantly outperforms the \RAG across all sizes of models. 
The improvement is especially notable under the \textit{Counterfactual Noise} setting, where the context is intentionally misleading and harder to distinguish, highlighting the effectiveness of our method in mitigating the influence of deceptive noise.
\textbf{2) Incorporating passages enhances robustness.} 
Compared to \RAG, the \Ins variant yields some robustness improvements, but falls short of \Both. 
This suggests that explicitly injecting passages into the reasoning process is more effective in helping the model resist noisy context.
\textbf{3) Smaller gains on distillation-based models.} 
The improvements from \Both are less pronounced on DeepSeek-R1-Distill-Qwen-32B compared to the Qwen3 series. 
This trend is consistent with observations in Section~\ref{subsection:RAG_Results}.

\subsection{Performance under Gold Passages} 
\begin{figure}[ht]
    \centering
    \includegraphics[width=0.9\linewidth]{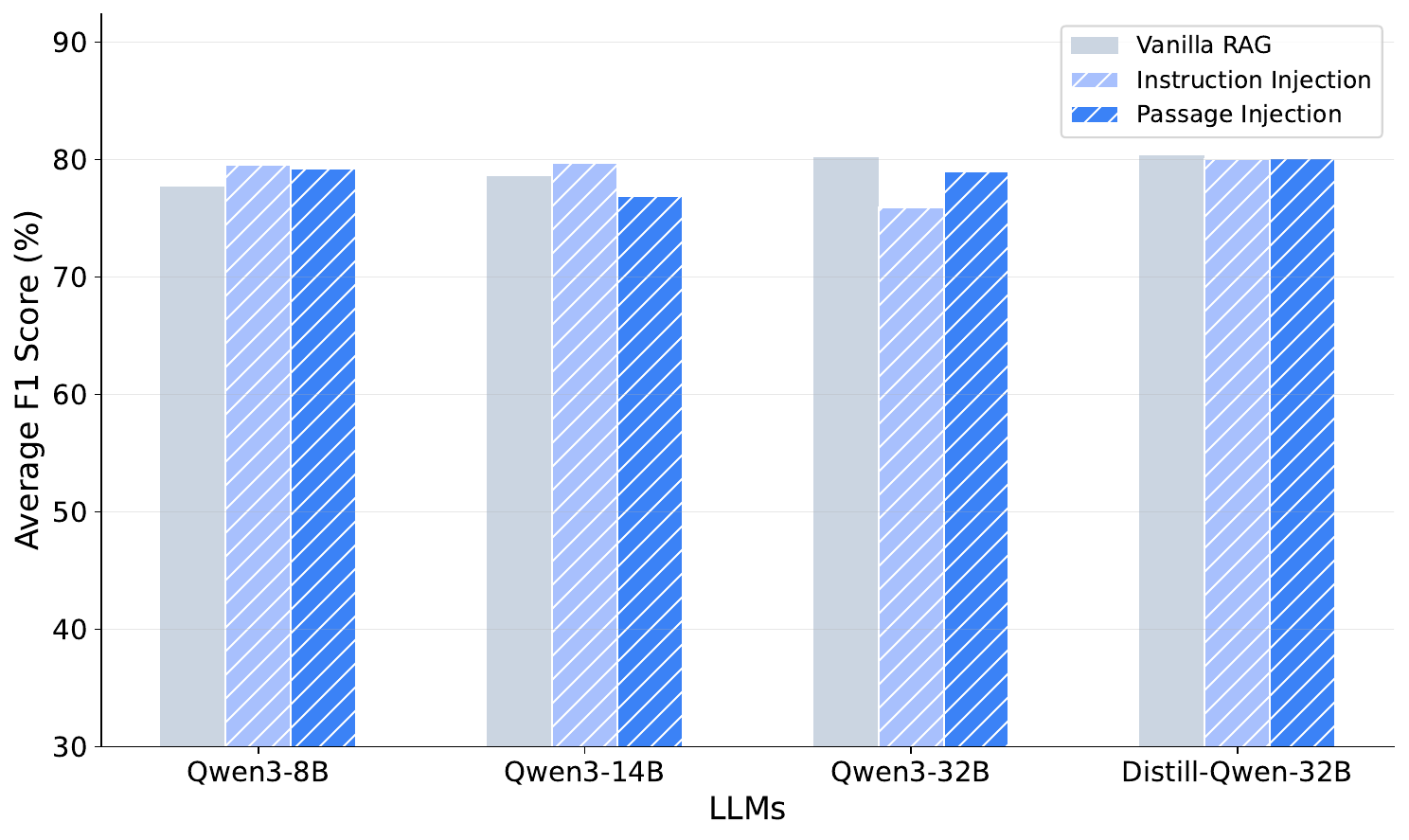}
    \caption{Average performance on 2WikiMultihopQA and HotpotQA using only gold passages. ``Distill-Qwen-32B'' refers to DeepSeek-R1-Distill-Qwen-32B.}
    \label{fig:Golden-Results}
\end{figure}

Figure~\ref{fig:Golden-Results} presents the results when only gold passages are provided.
In this setting, \Both performs comparably to \RAG, indicating that it improves robustness to noisy passages while still effectively leveraging helpful ones. 
This further suggests that the gains observed in general RAG settings primarily stem from enhanced robustness to noise.
Interestingly, \Both outperforms \RAG on Qwen3-8B, implying that smaller models may benefit more from having key information explicitly integrated into the reasoning process, whereas larger models can already extract relevant information directly from the input prompt.

\subsection{Analysis on Output Length}
Table~\ref{tab:Output_Len} presents the average output length (in characters) on CWQ and PopQA. 
Compared to \RAG, \Both significantly reduces the output length across both datasets and models.
This suggests that explicitly injecting passages into the reasoning process helps mitigate overthinking and encourages the model to produce more concise answers which is consistent with Section~\ref{sec:method}. \looseness=-1

\begin{table}[ht]
\centering
\caption{Average output length (in characters) of different RAG methods on CWQ and PopQA.}
\begin{tabular}{cccc}
    \toprule
    LLM	& RAG Method & CWQ	& PopQA \\
    \midrule
    
    \multirow{2}{*}{Qwen3-32B} & \RAG  & 2,267 & 1,760 \\
     & \Both  & 1,199 & 787 \\
     \midrule
     
    \multirow{2}{*}{\begin{tabular}[c]{@{}c@{}}DeepSeek-R1- \\ Distill-Qwen-32B\end{tabular}}	& \RAG & 1,909 & 1,408 \\
     & \Both & 1,190 & 774 \\
    \bottomrule
\end{tabular}
\label{tab:Output_Len}
\end{table}

\section{Conclusion}

In this work, inspired by reasoning-enhanced LLMs' ability to detect and revise reasoning errors, we propose \textbf{Passage Injection}—a method that explicitly integrates retrieved passages into the model’s reasoning process to enhance robustness against noisy information and improve overall RAG performance.
Experimental results on four reasoning-enhanced LLMs across four factual QA datasets demonstrate that Passage Injection significantly improves overall RAG performance.
Further analysis on two noisy retrieval settings-random noise, where the model is provided irrelevant passages, and counterfactual noise, where it is given misleading passages-shows that Passage Injection consistently improves LLMs' robustness to noisy passages.
Controlled experiments confirm that Passage Injection can also effectively leverage helpful passages. These findings suggest that incorporating passages in LLMs' reasoning process is a promising direction for building more robust RAG systems.

\begin{acks}
This work was funded by the National Natural Science Foundation of China (NSFC) under Grant No. 62302486, the Innovation Funding of ICT CAS under Grant No. E361140, the CAS Special Research Assistant Funding Project, the project under Grants No. JCKY2022130C039, the Strategic Priority Research Program of the CAS under Grants No. XDB0680102, and the NSFC Grant No. 62441229.
\end{acks}

\bibliographystyle{ACM-Reference-Format}

\balance

\bibliography{sample-base}
\end{document}